# The SCUBA-2 Cosmology Legacy Survey: 850$\mu$m maps, catalogues and number counts

J. E. Geach[1*], J. S. Dunlop[2], M. Halpern[3], Ian Smail[4], P. van der Werf[5], D. M. Alexander[4], O. Almaini[6], I. Aretxaga[7], V. Arumugam[2,8], V. Asboth[3], M. Banerji[9], J. Beanlands[10], P. N. Best[2], A. W. Blain[11], M. Birkinshaw[12], E. L. Chapin[13], S. C. Chapman[14], C-C. Chen[4], A. Chrysostomou[15], C. Clarke[16], D. L. Clements[17], C. Conselice[6], K. E. K. Coppin[1], W. I. Cowley[18], A. L. R. Danielson[4], S. Eales[19], A. C. Edge[4], D. Farrah[20], A. Gibb[3], C. M. Harrison[4], N. K. Hine[1], D. Hughes[7], R. J. Ivison[2,8], M. Jarvis[21,22], T. Jenness[3], S. F. Jones[23], A. Karim[24], M. Koprowski[1], K. K. Knudsen[23], C. G. Lacey[18], T. Mackenzie[3], G. Marsden[3], K. McAlpine[22], R. McMahon[8], R. Meijerink[5,25], M. J. Michałowski[3], S. J. Oliver[16], M. J. Page[26], J. A. Peacock[2], D. Rigopoulou[21,27], E. I. Robson[2,28], I. Roseboom[2], K. Rotermund[14], Douglas Scott[3], S. Serjeant[29], C. Simpson[30], J. M. Simpson[2], D. J. B. Smith[1], M. Spaans[25], F. Stanley[4], J. A. Stevens[1], A. M. Swinbank[4], T. Targett[31], A. P. Thomson[4], E. Valiante[19], T. M. A. Webb[32], C. Willott[33], J. A. Zavala[7], M. Zemcov[34,35]

[1] *Centre for Astrophysics Research, School of Physics, Astronomy & Mathematics, University of Hertfordshire, Hatfield, AL10 9AB*
*Other affiliations at end*

13 July 2016

**ABSTRACT**

We present a catalogue of nearly 3,000 submillimetre sources detected ($\geqslant 3.5\sigma$) at 850$\mu$m over $\sim$5 deg$^2$ surveyed as part of the James Clerk Maxwell Telescope (JCMT) SCUBA-2 Cosmology Legacy Survey (S2CLS). This is the largest survey of its kind at 850$\mu$m, probing a meaningful cosmic volume at the peak of star formation activity and increasing the sample size of submillimetre galaxies selected at 850$\mu$m by an order of magnitude. We describe the wide 850$\mu$m survey component of S2CLS, which covers the key extragalactic survey fields: UKIDSS-UDS, COSMOS, *Akari*-NEP, Extended Groth Strip, Lockman Hole North, SSA22 and GOODS-North. The average 1$\sigma$ depth of S2CLS is 1.2 mJy beam$^{-1}$, approaching the SCUBA-2 850$\mu$m confusion limit, which we determine to be $\sigma_c \approx 0.8$ mJy beam$^{-1}$. We measure the single dish 850$\mu$m number counts to unprecedented accuracy, reducing the Poisson errors on the differential counts to approximately 4% at $S_{850} \approx 3$ mJy. With several independent fields, we investigate field-to-field variance, finding that the number counts on 0.5–1° scales are generally within 50% of the S2CLS mean for $S_{850} > 3$ mJy, with scatter consistent with the Poisson and estimated cosmic variance uncertainties, although there is a marginal (2$\sigma$) density enhancement in the GOODS-North field. The observed number counts are in reasonable agreement with recent phenomenological and semi-analytic models, although robustly determining the shape of the faint end slope ($S_{850} < 3$ mJy) remains a key test. Finally, the large solid angle of S2CLS allows us to measure the bright-end counts: at $S_{850} > 10$ mJy there are approximately ten sources per square degree, and we detect the distinctive up-turn in the number counts indicative of the detection of local sources of 850$\mu$m emission, and strongly lensed high-redshift galaxies. Here we describe the data collection and reduction procedures and present calibrated maps and a catalogue of sources; these are made publicly available.

* E-mail: j.geach@herts.ac.uk

# 1 INTRODUCTION

Nearly a quarter of a century has passed since it was predicted that submillimetre observations could provide important insights into the nature of galaxies in the early Universe beyond the reach of optical and near-infrared surveys (Blain & Longair 1993). If early star-forming galaxies contained dust then ultraviolet photons should be reprocessed through the far-infrared (Hildebrand 1983) and redshifted into the submillimetre. Early observations certainly showed that some high-redshift sources are emitting a large fraction of their bolometric emission in the rest-frame far-infrared, detectable in the submillimetre, with integrated luminosities comparable to or exceeding local ultraluminous (ULIRG, $10^{12} L_\odot$) infrared galaxies (Rowan-Robinson et al. 1991; Clements et al. 1992). We now know that the far-infrared background (FIRB, Puget et al. 1996; Lagache et al. 1998; Fixsen et al. 1998) represents about half of the energy density associated with star formation integrated over the history of the Universe (Dole et al. 2006) and the peak of the volume averaged star formation rate density (SFRD) occurred at $z \sim 1$–3, to which submillimetre sources are expected to contribute significantly (Devlin et al. 2009). Identifying and characterising the galaxies contributing to the FIRB was (and remains) a major goal, and motivates blank-field submillimetre surveys.

About two decades ago the first submillimetre maps of the high-redshift Universe were made (Smail et al. 1997; Barger et al. 1998; Hughes et al. 1998; Lilly et al. 1999), opening a new window onto early galaxies. With twenty years of follow-up work across the electromagnetic spectrum we now have a good grasp of the nature of 'Submillimetre Galaxies' (SMGs) and their cosmological significance[1]. Nevertheless, the picture is far from complete. SMGs selected at 850$\mu$m lie at $\langle z \rangle \approx 2$–3 (e.g. Chapman et al. 2005, Pope et al. 2005; Wardlow et al. 2011; Simpson et al. 2014; Koprowski et al. 2014), are massive (Swinbank et al. 2004; Hainline et al. 2011; Michalowski et al. 2012), gas-rich (Greve et al. 2005; Tacconi et al. 2006, 2008; Engel et al. 2010; Carilli et al. 2010; Bothwell et al. 2013) and are associated with large supermassive black holes (Alexander et al. 2005, 2008; Wang et al. 2013). These properties make SMGs the obvious candidates for the progenitor population of massive elliptical galaxies today, seen at a time of rapid assembly a few billion years after the Big Bang (Lilly et al. 1999; Genzel et al. 2003; Swinbank et al. 2006), with star formation rates in the range 100–1000 $M_\odot$ yr$^{-1}$ derived from their integrated infrared luminosities (e.g. Magnelli et al. 2012; Swinbank et al. 2014).

The formation mechanism of SMGs remains in debate: by analogy with local ULIRGs, which are almost exclusively merging systems, it is predicted that SMGs form during major mergers of gas-dominated discs (Baugh et al. 2005; Ivison et al. 2012), triggering starbursts and central black hole growth. There is certainly observational evidence to support this, perhaps most convincingly in morphology and gas kinematics (e.g. Tacconi et al. 2010; Swinbank et al. 2010; Alaghband-Zadeh et al. 2012; Chen et al. 2015). On the other hand, hydrodynamic simulations may be able to reproduce the properties of SMGs without the need for mergers, for example if there is a prolonged ($\sim 1$ Gyr) phase of gas accretion which drives high star formation rates, where cooling is accelerated through metal enrichment at early times (e.g. Narayanan et al. 2015, see also Davé et al. 2010). In recent semi-analytic models, starbursts triggered by bar instabilities in galaxy discs are the dominant mechanism producing SMGs in model universes (Lacey et al. 2015), and indeed there is some empirical evidence that SMGs have optical/near-infrared morpholigies consistent with discs (e.g. Targett et al. 2013).

Observations in the 850$\mu$m atmospheric window offer a unique probe of the distant Universe, owing to the so-called 'negative $k$-correction' (Blain & Longair 1993). For cosmological sources, the 850$\mu$m band probes the Rayleigh-Jeans tail of the cold dust continuum emission of carbonaceous and silicate grains in thermal equilibrium in the stellar ultraviolet radiation field. As the thermal spectrum is redshifted, cosmological dimming is compensated for by increasing power as one 'climbs' the Rayleigh-Jeans tail as it is redshifted through the band. Thus, two sources of equal luminosity will be observed with roughly the same flux density at 850$\mu$m at $z \approx 0.5$ and $z \approx 10$. As a guide, a galaxy in the ultraluminous class (with $L_{\rm IR} \approx 10^{12} L_\odot$) is observed with a flux density of 1–2 mJy at 850$\mu$m over most of cosmic history (Blain et al. 2002). For this reason, flux limited surveys at 850$\mu$m offer the opportunity to sample huge cosmic volumes, potentially probing well into the epoch of re-ionization.

Despite the large redshift depth probed by deep 850$\mu$m surveys, the solid angle subtended by existing surveys, and their sensitivity, has been bounded by technology: until recently, submillimetre cameras have been limited in field-of-view and sensitivity that has made degree-scale mapping difficult. However, submillimetre imaging technology has blossomed over the past twenty years. At first only single channel broadband submillimeter photometers were available in (e.g. Duncan et al. 1990), making survey work impossible. Then the first cameras came online, mounted on 10–15 m single dish telescopes such as the Caltech Submillimeter Observatory (CSO) and the James Clerk Maxwell Telescope (JCMT): the Submillimeter High Angular Resolution Camera (SHARC, Wang et al. 1996) and the Submillimetre Common-User Bolometer Array (SCUBA, Holland et al. 1999) using small arrays of 10s of bolometers covering just a few arcminutes field of view. These arrays enabled the first extragalactic submillimetre surveys (Smail et al. 1997; Hughes et al. 1998), but covering a cosmologically representative solid angle at the necessary depth was still tremendously expensive in terms of observing time.

Further cameras based on bolometer arrays were developed through the late 1990s: Bolocam (Glenn et al. 1998), MAMBO (Kreysa et al. 1998), SHARC-II (Dowell et al. 2002), LABOCA (Siringo et al. 2009) and AzTEC (Wilson et al. 2008) and the scale of extragalactic submillimetre surveys grew in tandem (e.g. Eales et al. 2000; Scott et al. 2002, 2006; Borys et al. 2003; Webb et al. 2003; Greve et al. 2004; Coppin et al. 2006; Weiss et al. 2009; Scott et al. 2010, 2012, Austermann et al. 2010). Unfortunately the semiconductor technology underlying the first and second generation of submillimetre cameras is not scalable, limiting bolometer arrays to around 100 pixels. A solution was found in superconducting transistion edge sensors (TES, see Irwin et al. 1995) coupled with Superconducting Quantum Interference Device (SQUID) amplifiers, that allowed for the construction of submillimetre sensitive bolometer arrays an order of magnitude larger than previously achieved. Clearly this opened up the possibility of performing much larger, more efficient submillimetre surveys than had ever been possible before from the ground.

---

[1] It is worth noting that it is now common to refer to SMGs as cosmological sources selected right across the 250–1000$\mu$m wavelength range. With the high-altitude Balloon-borne Large Aperture Submillimeter Telescope (BLAST, Pascale et al. 2008) and then the launch of the *Herschel Space Observatory* in 2009 (Griffin et al. 2008) the path has been opened up to large area submillimetre surveys at $\lambda \lesssim 650\mu$m (e.g. Eales et al. 2010), although suffering from high confusion noise due to the limited size of dishes that can be flown in the sky and space.

The second generation SCUBA camera, SCUBA-2, on the JCMT is the first of such large format instruments using TES technology (Holland et al. 2013). SCUBA-2 comprises two arrays (for the 450$\mu$m and 850$\mu$m bands) of 5120 bolometers each, covering an 8 arcminute field of view. With mapping speeds (to equivalent depth) over an order of magnitude faster than its predecessor, SCUBA-2 has enabled a huge leap in submillimetre survey science. TES focal plane arrays have also formed the basis of other recent submillimetre instrumentation, such as the South Pole Telescope (Carlstrom et al. 2011) and Atacama Cosmology Telescope (Swetz et al. 2011). Future large format submillimetre cameras are likely to make increasing use of Kinetic Inductance Detectors (KIDS, Day et al. 2003): the New Instrument of KIDS Arrays (NIKA2) on the 30 metre Institut de Radioastronomie Millimétrique (IRAM) telescope (Monfardini et al. 2010) uses this new detector technology.

Soon after commissioning of SCUBA-2, five JCMT 'Legacy Surveys' (JLS) commenced. The largest of these is the JCMT SCUBA-2 Cosmology Legacy Survey (S2CLS). In this paper we present the wide 850$\mu$m survey component of the S2CLS, presenting maps and a source catalogue for public use. This paper is organised as follows: in §2 we define the survey and describe data reduction and cataloguing procedures; in section §3 we present the maps and catalogues and in §4 we use these data to measure the number counts of 850$\mu$m-selected sources with the best statistical precision to date, including an analysis of the impact of cosmic variance on scales of $\sim$1 degree. We summarize the paper in §5. Where relevant, we adopt a fiducial $\Lambda$CDM cosmology with $\Omega_{\rm m}=0.3$, $\Omega_\Lambda = 0.7$ and $H_0 = 70\,{\rm km\,s^{-1}\,Mpc^{-1}}$.

## 2 THE SCUBA-2 COSMOLOGY LEGACY SURVEY

The S2CLS survey has two tiers: wide and deep. The wide tier covers several well-explored extragalactic survey fields: *Akari*-Northern Ecliptic Pole, COSMOS, Extended Groth Strip, GOODS-North, Lockman Hole North, SSA22 and UKIDSS-UDS (Figure 1, Table 1), mapping at 850$\mu$m during conditions where the zenith optical depth at 225 GHz was $0.05 < \tau_{225} \leqslant 0.1$ and field elevations exceeded 30 degrees. In the deep tier several deep 'keyhole' regions within the wide fields were mapped when $\tau_{225} \leqslant 0.05$, conditions suitable for obtaining 450$\mu$m maps which require the lowest opacities (Geach et al. 2013). Note that SCUBA-2 simultaneously records 450$\mu$m and 850$\mu$m photons, and while the complementary 450$\mu$m data exist for the wide 850$\mu$m maps we present here, they have not been processed, since they are not expected to be of sufficient quality given the observing conditions. In this paper we present the maps (Figure 1) and catalogue from the wide tier only.

### 2.1 Observations

The S2CLS was conducted for just over three years, from December 2011 to February 2015; Figure 2 shows the time distribution of observations during the survey. The wide tier used the PONG mapping strategy for large fields, whereby the array is slewed around the target (map centre) in a path that 'bounces' off the rectangular edge of the defined map area in a manner reminiscent of the classic arcade game (Thomas et al. 2014). The PONG pattern ensures that the array makes multiple passes back and forth between the map extremes, filling the square mapping area. To ensure uniform coverage the field is rotated 10–15 times (depending on map size) during an observation, resulting in a circular field with uniform sensitivity over the nominal mapping area (but with science-usable area beyond this, see §2.4.1). Scanning speeds were 280$''$ sec$^{-1}$ for maps of size 900$''$ up to 600$''$ sec$^{-1}$ for the largest single map of 3300$''$. Observations were limited to 30–40 minutes each to monitor variations in observing conditions, with regular pointing calibrations performed throughout the night. Typical pointing corrections are of order $\sim$1$''$ between observations. In addition to the zenithal opacity constraints described above, elevation constraints were also imposed: to ensure sufficiently low airmass, targets were only observed when above 30 degrees, and a maximum elevation constraint of 70 degrees was also imposed (only relevant for the COSMOS field). This high elevation constraint was set because it was found that the telescope could not keep pace with the alt-az demands of the scanning pattern, resulting in detrimental artifacts in the maps. Since the Lockman Hole North field is observable during COSMOS transit, the strategy was simply to switch targets as COSMOS rose above 70 degrees.

For all but the EGS and COSMOS field, the targets were mapped with single PONG scans with diameters ranging from 900–3300$''$ (Table 1). The EGS was mapped using a chain of six 900$''$ PONG maps (each slightly overlapping) to optimise coverage of the multiwavelength data along the multiwavelength strip. In COSMOS the mapping strategy was a mosaic consisting of a central 900$''$ PONG and four 2700$''$ PONG maps offset by 1147$''$ in RA and Dec. from the central map, forming a 2 $\times$ 2 grid of 'petals' around the central PONG, with some overlap. This was deemed preferable to obtaining a single very large PONG map encompassing the full field, allowing depth to be built-up in each tile sequentially. Only $\sim$50 per cent of the COSMOS area was completed to full depth, due to the end of JCMT operations by the original partners. The full 2 $\times$ 2-degree field is now being completed as part of a follow-on project 'S2-COSMOS' (PI: Ian Smail and J. Simpson et al. 2016 in preparation). Figure 1 shows a montage of the S2CLS fields to scale, and Figure 3 shows an example of the sensitivity variation across a single PONG map (the UKIDSS-UDS field), illustrating the homogeneity of the noise coverage across the bulk of the scan region, with instrumental noise varying by just $\sim$5% across degree scales. We describe the process to create the S2CLS 850$\mu$m maps in the following section.

### 2.2 Data reduction

Each SCUBA-2 bolometer records a timestream, where the signal is a contribution of background (mainly sky and ambient emission), astronomical signal and noise. The basic principle of the data reduction is to extract astronomical signal from these timestreams and map them onto a two dimensional celestial projection. We have used the Dynamical Iterative Map-Maker (DIMM) within the *Sub-Millimetre Common User Reduction Facility* (SMURF; Chapin et al. 2013). We refer readers to Chapin et al. (2013) for a detailed overview of SMURF, but describe the main steps, including specific parameters we have chosen for the reduction of the blank-field maps, here (see also Geach et al. 2013).

First, time-streams are downsampled to a rate matching the pixel scale of the final map, based on the scanning speed (§2.1). All S2CLS maps are projected on a tangential co-ordinate system with 2$''$ pixels. Flat-fields are then applied to the time-streams using flat scans that bracket each observation, and a polynomial baseline fit is subtracted from each bolometer's time-stream (we actually use a linear – i.e. order 1 – fit). Then each time-stream is cleaned for spikes (using a 5$\sigma$ threshold in a box size of 50 samples), DC steps are removed and gaps filled. After cleaning, the DIMM enters an iterative process that aims to fit the data with a model comprising a common-mode fluctuating atmospheric signal, positive astronomical signal and instrumental and fine-scale atmospheric noise. The common mode modelling is performed independently for each SCUBA-2 sub-array, deriving a template for the average signal seen by all the bolometers. The common mode is then removed, and an extinction correction is applied (Dempsey et al. 2013). Next, a filtering step is performed in the Fourier domain, which rejects power at frequencies corresponding to angular scales $\theta > 150''$ and $\theta < 4''$. The next step is to estimate the astrononmical signal. This is done by gridding the time-streams onto the celestial projection; since each pixel will be sampled many times by independent bolometers (slewing over the sky in the PONG scanning pattern), then the positive signal in a given pixel can be taken to be an accurate estimate of the astronomical signal (assuming the previous steps have eliminated all other sources of emission or spikes, etc.). This model of the astronomical signal is then projected *back* to a time-stream and subtracted from the data. Finally, a noise model is estimated for each bolometer by measuring the residual, which is then used to weight the data during the mapping process in additional steps. The iterative process above runs until convergence is met. In this case, we execute a maximum of 20 iterations, or terminate the process when the map tolerance $\Delta\chi^2$ reaches 0.05.

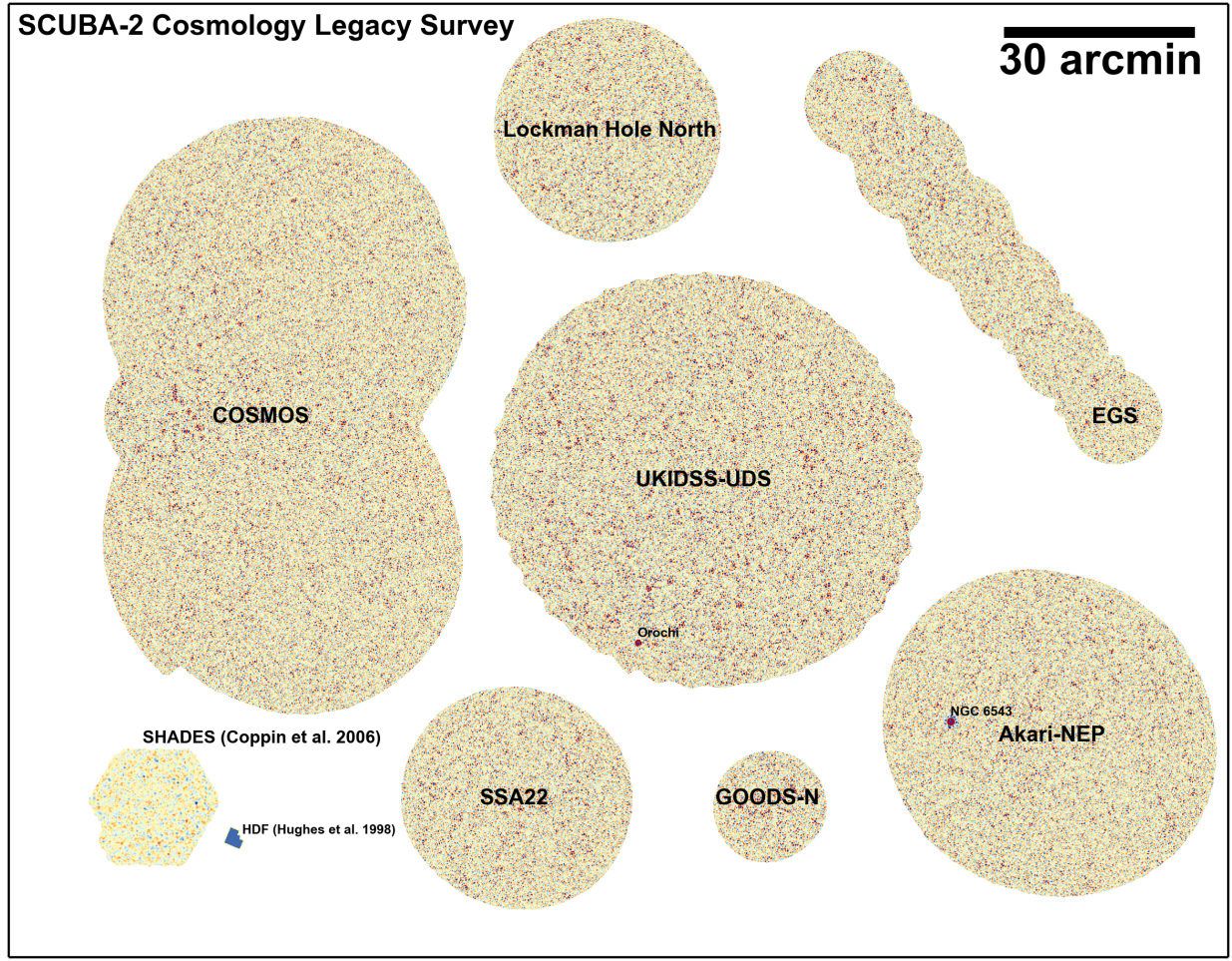

**Figure 1.** The JCMT SCUBA-2 Cosmology Legacy Survey: montage of signal-to-noise ratio maps indicating relative coverage in the seven extragalactic fields (see also Table 1). This survey has detected approximately 3,000 submillimetre sources over approximately 5 square degrees. The two bright sources identified are 'Orochi', an extremely bright SMG first reported by Ikarashi et al. (2010) in UKIDSS-UDS, and NCG 6543 in *Akari*-NEP. For scale comparison we show the 850$\mu$m map of the UKIDSS-UDS from the SCUBA HAlf DEgree Survey (SHADES, Coppin et al. 2006) and the footprint of the *Hubble Space Telescope* WFPC2, corresponding to the size of the SCUBA map of the *Hubble* Deep Field from Hughes et al. (1998) – one of the first deep extragalactic maps at 850$\mu$m. Note that the size of the primary beam of the Atacama Large Millimeter/submillimeter Array (ALMA) at 850$\mu$m is comparable to the size of the JCMT beam: the full S2CLS survey subtends a solid angle over 100,000 times the ALMA primary beam at 850$\mu$m. The angular scale of 30′ subtends approximately 5 co-moving Mpc at the typical redshift of the SMG population, $z \approx 2$.

**Table 1.** S2CLS survey fields (see also Figure 1). Right Ascension and Declination refer to the central pointing (J2000). The area corresponds to map regions where the root mean squared instrumental noise is below 2 mJy. Note that at the end of the survey, the COSMOS field was only 50% completed; remainder is now being observed to equivalent depth in a new survey (S2-COSMOS, PI: Smail; J. Simpson et al. in preparation).

| Field name | R.A. | Dec. | Area (deg$^2$) | 1$\sigma$ 850$\mu$m depth (mJy beam$^{-1}$) | Scan recipe | Astrometric reference |
|---|---|---|---|---|---|---|
| *Akari*-North Ecliptic Pole | 17 55 53 | +66 35 58 | 0.60 | 1.2 | 45′ PONG | Takagi et al. (2012) 24$\mu$m |
| COSMOS | 10 00 30 | +02 15 02 | 2.22 | 1.6 | 2×2 45′ PONG | Sanders et al. (2008) 3.6$\mu$m |
| Extended Groth Strip | 14 17 41 | +52 32 15 | 0.32 | 1.2 | 6×1 15′ PONG | Barmby et al. (2008) 3.6$\mu$m |
| GOODS-N | 12 36 51 | +62 12 52 | 0.07 | 1.1 | 15′ PONG | *Spitzer*-GOODS-N MIPS 24$\mu$m catalogue[†] |
| Lockman Hole North | 10 46 07 | +59 01 17 | 0.28 | 1.1 | 30′ PONG | Surace et al. (2005) 3.6$\mu$m |
| SSA22 | 22 17 36 | +00 19 23 | 0.28 | 1.2 | 30′ PONG | Lehmer et al. (2009) 3.6$\mu$m |
| UKIDSS-Ultra Deep Survey | 02 17 49 | −05 05 55 | 0.96 | 0.9 | 60′ PONG | UKIDSS-UDS Data Release 8 3.6$\mu$m[*] |

[†]irsa.ipac.caltech.edu/data/SPITZER/docs/spitzermission/observingprograms/legacy/goods
[*]www.nottingham.ac.uk/astronomy/UDS

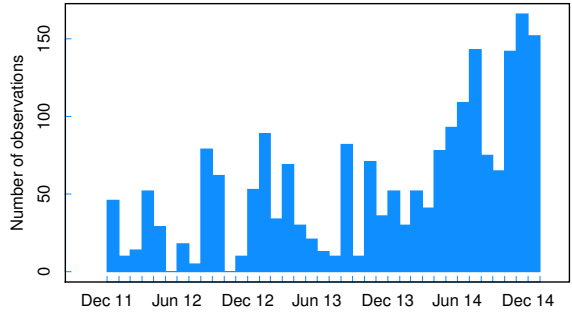

**Figure 2.** Time distribution of 850$\mu$m observations. In total CLS conducted 2041 wide-field observations on 320 nights from November 2011 to Febuary 2015. The increase in frequency of observations towards the end of the survey reflects the effect of 'extended observing' into the post-sunrise morning hours when the opacity and conditions were still suitable for observations. Note that one observation is equivalent to 30-40 minutes of integration time.

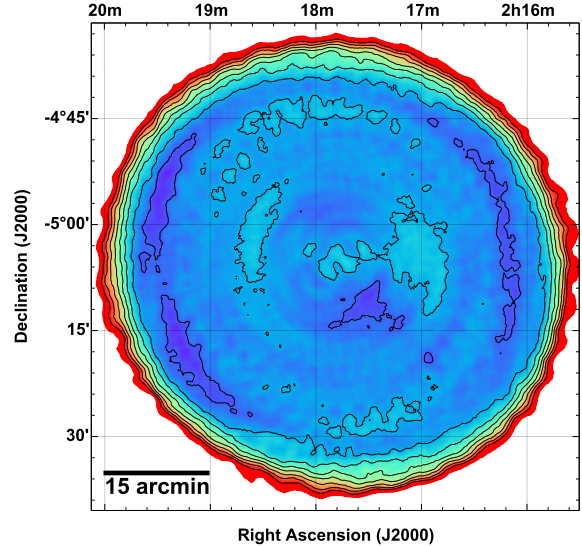

**Figure 3.** An example of the sensitivity coverage in a single S2CLS field. This map shows the instrumental noise map of the UKIDSS-UDS (a single PONG), scaled between $\sigma_{\rm instr} = 0.8$–$1.2$ mJy. Contours are at steps of 0.05 mJy starting at 0.8 mJy. This demonstrates the uniform nature of the PONG map over the majority of the mapping region, radially rising beyond the nominal extent of the area scanned to uniform depth (effectively over-scan regions receiving shorter integration time).

S2CLS obtained many individual scans of each field. The DIMM allows for all the scans to be simultaneously reduced in the manner described above. However, we adopt an approach where the DIMM is only given individual observations, producing a set of maps for each target field which can then be co-added into a final stack. For this we use the PICARD recipe *mosaic_jcmt_images* which uses the WCSMOSAIC task within the STARLINK KAPPA package, weighting each input image by the inverse variance per pixel. With a set of individual observations for each field we can also construct maps of sub-sets of the data and produce jackknife maps where a random 50% of the images are inverted, thus removing astronomical signal in the final stack, and generating source-free noise realisations of each field (Figure 4); useful for certain statistical tests.

The last processing step is to apply a matched filter to the maps, convolving with the instrumental PSF to optimize the detection of point sources. We use the PICARD recipe *scuba2_matched_filter* which first smooths the map (and the PSF) with a 30$''$ Gaussian kernel, then subtracts this from both to remove any large scale structure not eliminated in the filtering steps that occurred during the DIMM reduction. The map is then convolved with the smoothed beam. A flux conversion factor of 591 Jy beam$^{-1}$ pW$^{-1}$ is applied to give the maps units of flux density. This canonical calibration is the average value derived from observations of hundreds of standard submillimetre calibrators observed during the S2CLS campaign (Dempsey et al. 2013). The filtering steps employed in the data reduction, including the match-filtering step, introduce a slight (10%) loss of response to point sources. We have measured this loss by injecting a model source of known (bright) flux density into the data and recovering its flux after filtering; we correct for this in the flux calibration. The absolute flux calibration is expected to be accurate to within 15%.

### 2.3 Astrometric refinement and registration

The JCMT pointing is regularly checked against standard calibrators during observations, with typical pointing drift corrections typically of order 1–2$''$; similar to the pixel scale at which the maps are gridded. To improve the astrometric refinement of the final co-added maps we adopt a maximal signal-to-noise stacking technique: for each field we use a mid-infrared selected catalogue and stack the submillimetre maps at the positions of reference sources to measure a high-significance statistical detection. We repeat the process many times, updating the world coordinate system reference pixel coordinates at each step with small $\Delta\alpha$ and $\Delta\delta$ increments. The goal is to find the ($\Delta\alpha$, $\Delta\delta$) that maximise the signal-to-noise of the stack in the central pixel. We iterate over several levels of refinement until no further change in ($\Delta\alpha$, $\Delta\delta$) is required. The average changes to the astrometric solution are of order 1–2$''$, comparable to the pixel scale and similar to the source positional

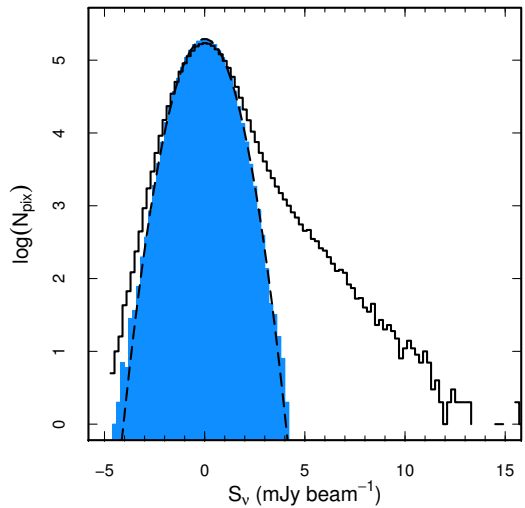

**Figure 4.** Distribution of pixel values in the UKIDSS-UDS flux density map, showing the characteristic tail representing astronomical emission. The shaded region shows the equivalent distribution in a jackknife map, constructed by inverting a random half of the data before co-addition. The dashed line is simply a normal distribution with zero mean and scale set to the standard deviation of pixel values in the jackknife map, illustrating that the noise in the map is approximately Gaussian.

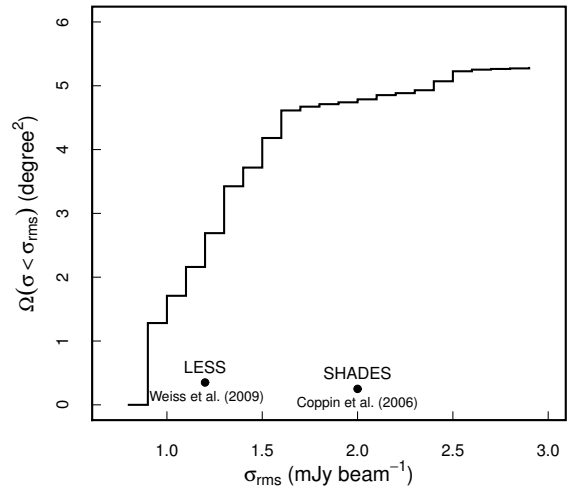

**Figure 5.** Cumulative area of the SCUBA-2 Cosmology Legacy Survey as a function of sensitivity, compared to the largest previous 850$\mu$m surveys SHADES (Coppin et al. 2006) and (at 870$\mu$m) LESS (Weiss et al. 2009). The majority of S2CLS reaches a sensitivity of below 2 mJy beam$^{-1}$, a dramatic step forward compared to previous surveys in the same waveband.

uncertainty (see §2.5.2). Table 1 lists the reference catalogues used for each field.

### 2.4 Statistics

#### 2.4.1 Area coverage

The PONG scanning strategy results in maps that are uniformly deep over the nominal scanning area, however the usable area in each map is larger than this because of overscan, with radially increasing noise due to the lower effective exposure time in these regions. Although shallower than the map centres, these annular regions around the perimeters of the fields, are deep enough to detect sources. Figure 5 shows the cumulative area of the survey as a function of (instrumental) noise. The total survey area is approximately 5 square degrees, with >90% of the survey area reaching a sensitivity of under 2 mJy beam$^{-1}$.

#### 2.4.2 Modelling the PSF

The matched-filtering step described in §2.2 modifies the shape of the instrumental PSF, effectively slightly broadening it and increasing the depth of bowling. We derive an empirical PSF by stacking 322 >5$\sigma$ significance point sources in the UKIDSS-UDS map and fit an analytic surface function to the average profile. The profile is shown in Figure 6 in comparison to the instrumental PSF, and has a FWHM of 14.8″. Two-dimensional fitting of the stack reveals that the beam profile $P(\theta)$ is circular to within 1% and can be fit with the superposition of two Gaussian functions:

$$P(\theta) = A\exp\left(\frac{\theta^2}{2\sigma^2}\right) - 0.98 A\exp\left(\frac{\theta^2}{2.04\sigma^2}\right) \quad (1)$$

with $A = 41.4$ and $\sigma = 9.6''$.

#### 2.4.3 The confusion limit

The confusion limit (Scheuer 1957) $\sigma_c$ is the flux level at which the pixel-to-pixel variance $\sigma^2$ no longer reduces with exposure time due to crowding of the beam by faint sources. The total variance is a combination of the instrumental noise $\sigma_i$ (in units of mJy beam$^{-1}\sqrt{s}$) and the confusion noise (in units of mJy beam$^{-1}$):

$$\sigma^2 = \sigma_i^2 t^{-1} + \sigma_c^2. \quad (2)$$

We can evaluate the confusion limit by measuring $\sigma^2$ directly from the pixel data in a progression of maps as we sequentially co-add new scans. Figure 7 shows how the variance evolves as a function of inverse *pixel* integration time for the central 15′ of the UKIDSS-UDS, which reaches an instrumental noise of 0.8 mJy beam$^{-1}$. The best fit $\sigma_c$ is 0.8 mJy beam$^{-1}$; this confusion noise should be added in quadrature to instrumental and deboosting (§2.5.1) uncertainties when considering the flux density of sources.

### 2.5 Source extraction

The matched-filtering step optimises the maps for the detection of point sources – i.e. emission features identical to the PSF. To extract and catalogue sources we employ a simple top-down peak-finding algorithm: starting from the most significant peak in the signal-to-noise ratio map, the peak flux, noise and position of a source is catalogued before the source is removed from the flux (and signal-to-noise) map by subtracting a scaled version of the model PSF. The highest peak in the source-subtracted map is then catalogued and subtracted and so-on until a floor threshold significance is reached, below which 'detections' are no longer trusted. Note that this procedure can potentially deblend sources with markedly different fluxes. The floor detection limit is set to $3\sigma$ which allows us to explore the properties of the lowest-significance detections, noting that further cutting can be performed directly on the catalogue. In the following we assume a cut of $3.5\sigma$ as the formal detection limit of S2CLS, where we estimate that the false detection rate is approximately 20% (see §2.5.3).

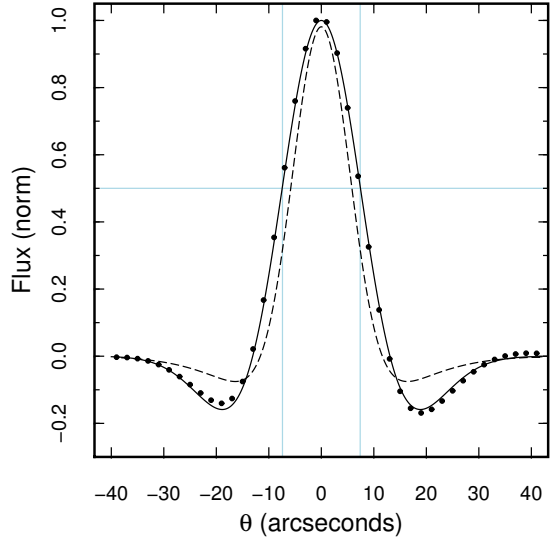

**Figure 6.** Model of the SCUBA-2 PSF. The dashed line shows the instrumental PSF (Dempsey et al. 2013), and the points show the shape of the average point source in the UKIDSS-UDS field, derived by stacking all sources detected at $5\sigma$ significance or greater. The maps are match-filtered, which includes a smoothing step that slightly broadens the instrumental PSF and deepens 'ringing'. The empirical PSF is well modelled with the superposition of two Gaussians (§2.3.2), is circular, and has a FWHM of $14.8''$.

### 2.5.1 Completeness and flux boosting

To evaluate source detection completeness we insert fake sources matching a realistic number count distribution into the jackknife noise maps of each field and then try to recover them using the source detection algorithm described above. We adopt the differential number counts fit of Casey et al. (2013) as a fiducial model, which has the Schechter form:

$$\frac{dN}{dS} = \left(\frac{N_0}{S_0}\right)\left(\frac{S}{S_0}\right)^{-\gamma} \exp\left(-\frac{S}{S_0}\right) \quad (3)$$

with $N_0 = 3300\,\mathrm{deg}^{-2}$, $S_0 = 3.7\,\mathrm{mJy}$ and $\gamma = 1.4$. We insert sources down to a flux density limit of 1 mJy and each source is placed at a random position into each map (we do not encode any clustering of the injected sources). An injected source is recovered if a point source is found above the detection threshold within $1.5\times$FWHM of the input position. This is a somewhat arbitrary, but generous, threshold, and if there are multiple injected sources within this radius, then we take the closest match. This procedure is repeated 5,000 times for each map, generating a set of mock catalogues containing millions of sources with a realistic flux distribution, allowing us to assess the completeness and flux boosting statistics.

The ratio of recovered sources to total number of input sources is evaluated in bins of input flux density and local (instrumental) noise. When applying completeness corrections we use the binned values as a look-up table, using two dimensional spline interpolation to estimate the completeness rate for a given source. Figure 8 compares the average completeness of each field (i.e. at the average depth of each map) as a function of intrinsic flux density. Table 2 lists the average 50% and 80% completeness limits for each field and the number of sources above each limit.

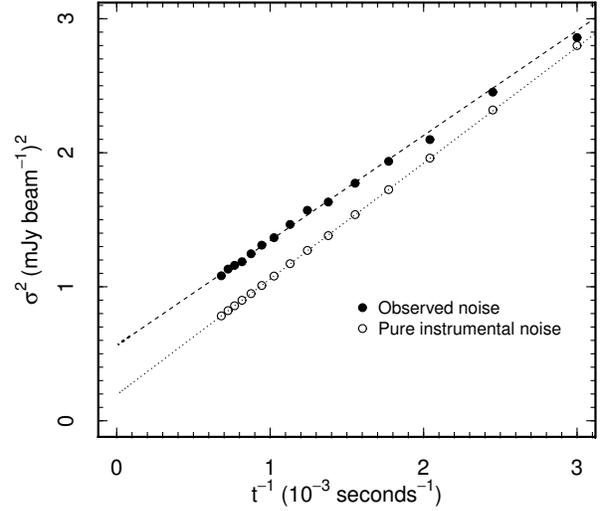

**Figure 7.** Measurement of the $850\mu$m confusion limit for SCUBA-2: we progressively co-add single exposures of the UKIDSS-UDS field, measuring the pixel-to-pixel root mean square value in the uniform central $15'$ of the beam-convolved flux map, whilst also tracking the fall off in the pure instrumental noise estimate. At infinite exposure the instrumental noise is projected to reach zero, whereas the non-zero intercept of the observed flux r.m.s. is the confusion limit (Equation 2). We measure this to be $\sigma_c \approx 0.8\,\mathrm{mJy\,beam}^{-1}$ averaged over the field. Note that the exposure time is the average per $2''$ pixel.

**Table 2.** 50% and 80% completeness limits for the S2CLS fields, quoted at the median map depth (Table 1). We also present the number of sources brighter than the 50% and 80% limits in each field ($N_{50,80}$). Note that these flux densities refer to the *deboosted* – i.e. intrinsic – flux densities. At the $5\sigma$ level observed flux densities are typically overestimated by 20% (§2.5.1).

| Field | 50% (mJy) | 80% (mJy) | $N_{50}$ | $N_{80}$ |
| --- | --- | --- | --- | --- |
| *Akari*-NEP | 4.1 | 5.2 | 132 | 59 |
| UKIDSS-UDS | 3.0 | 3.8 | 543 | 302 |
| COSMOS | 4.9 | 6.2 | 302 | 181 |
| Lockman Hole North | 3.6 | 4.6 | 96 | 49 |
| GOODS-N | 3.9 | 4.7 | 32 | 21 |
| Extended Groth Strip | 3.9 | 5.0 | 99 | 51 |
| SSA22 | 3.9 | 4.9 | 78 | 38 |

We can simultaneously evaluate flux boosting as a function of local noise and *observed* flux density simply by comparing the recovered flux to the input flux density of each source. Flux boosting is the overestimation of source flux when measurements are made in the presence of noise and is related to both Eddington and

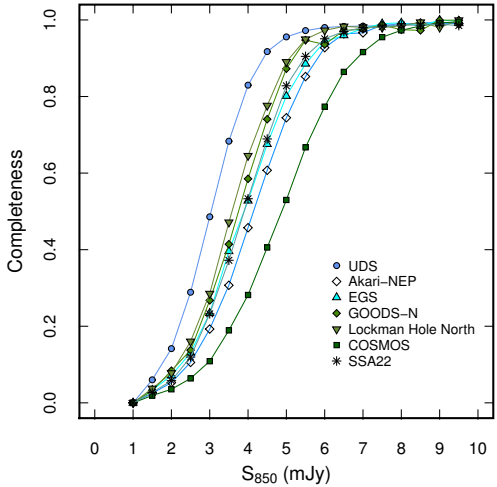

**Figure 8.** Completeness of the different S2CLS fields, derived from the recovery rate of fake sources injected into jackknife maps as a function of input flux, where a successful recovery at a detection significance of $3.5\sigma$. Note that the completeness falls to zero at 1 mJy as this corresponds to the limit of the injected source model; in practice it is possible that sub-mJy sources could be boosted above the detection limit. The 50% and 80% limits of each field are listed in Table 2.

Malmquist bias. Due to the statistical nature of boosting, a source with some observed flux density $S_{\rm obs}$ is actually drawn from a distribution of true flux density, $p(S_{\rm true})$. Our recovery procedure allows us to estimate $p(S_{\rm true})$, since we can simply measure the histogram of the injected flux density of sources in bins of $(S_{\rm obs}, \sigma)$. This method can be compared to the traditional Bayesian technique to estimate boosting (e.g. Jauncey et al. 1968; Coppin et al. 2005), such that the posterior probability distribution for an observed flux density can be expressed:

$$p(S_{\rm true}|S_{\rm obs}, \sigma) = \frac{p(S_{\rm true}) p(S_{\rm obs}, \sigma | S_{\rm true})}{p(S_{\rm obs}, \sigma)}. \qquad (4)$$

The likelihood of the data is given by assuming a Gaussian photometric error on the observed flux density, and the prior is simply the same assumed number counts model used in the simulations described above. Figure 9 compares the empirically-estimated $p(S_{\rm true})$ and the posterior probability distribution for $S_{\rm true}$ from Equation (4). The empirical distributions are truncated at 1 mJy because this is the faint limit of the injected source distribution; clearly we can not track individual sources fainter than this. An identical counts model is used as a prior in the Bayesian approach, but note that the posterior flux density distribution *does* extend below 1 mJy; this is because it is effectively the product of a Gaussian (the observed flux density and instrumental uncertainty) and the histogram of pixel values in a map of sources drawn from the model number counts, convolved with the beam. The two methods return similar results, although the empirical method systematically predicts a slightly smaller boosting factor $\mathcal{B} = S_{\rm obs}/S_{\rm true}$ than the Bayesian approach, with the two methods converging as $S_{\rm obs}$ increases. Note that neither method assumes any clustering of sources, which could well be important (Hodge et al. 2013; Simpson et al. 2015).

There are two important differences in the deboosting methods that may explain this: (i) the Bayesian approach does not consider noise (aside from the confusion noise arising from convolving the fake map with the beam), and, related, (ii) the posterior flux distribution derived in Equation 4 is not necessarily measured 'at peak', i.e. does not consider that the recovered position of a source can shift due to the presence of noise; in the empirical method, we account for such shifts. This relates to the 'bias-to-peak' discussed by Austermann et al. (2010). We adopt the 'empirical' approach in this work to deboost observed fluxes: we draw samples from the distribution of $S_{\rm true}$ for a given $(S_{\rm obs}, \sigma)$ and calculate the mean and variance of these true fluxes, with the latter providing the uncertainty on the deboosted flux density (provided in the source catalogue). We summarise the empirically derived completeness and boosting for each field, visualised in the plane of flux density and local instrumental noise, in Figure 10. In Figure 11 we show the average flux boosting as a function of signal-to-noise ratio in each field, indicating that at fixed detection significance, the level of flux boosting is consistent across the survey, with observed flux densities approximately 20% higher on average than the intrinsic flux density at the $5\sigma$ level. The average boosting is well described by a power law:

$$\mathcal{B} = 1 + 0.2 \left(\frac{\rm SNR}{5}\right)^{-2.3} \qquad (5)$$

### 2.5.2 Positional uncertainty

The simulations described above allow us to investigate the scatter in the difference between input position and recovered position. Like the completeness and boosting, we evaluate the average $\delta\theta$ between input and recovered position in bins of input flux density and local instrumental noise. Following Condon et al. (2007) and Ivison et al. (2007), for a given (Gaussian-like) beam, the positional accuracy is expected to scale with signal-to-noise. Figure 12 shows the mean difference between input and recovered source position as a function of signal-to-noise ratio for each field. We find that the positional uncertainty of S2CLS sources is well described by a simple power law, reminiscent of Equation B22 of Ivison et al. (2007):

$$\delta\theta = 1.2'' \times \left(\frac{\rm SNR}{5}\right)^{-1.6} \qquad (6)$$

### 2.5.3 False detection rate

To measure the false detection rate we compare the number of 'detections' in the jackknife maps to those in the real maps as a function of signal-to-noise ratio. By construction, the jackknife maps contain no astronomical signal and have Gaussian noise properties (Figure 4); therefore any detections are due to statistical fluctuations expected from Gaussian noise at the $\geqslant 3.5\sigma$ level. Figure 13 shows the false detection rate as a function signal-to-noise ratio; At our $3.5\sigma$ limit the false detection (or contamination) rate is 20%, falling to 6% at $4\sigma$ and falls below 1% for a $\geqslant 5\sigma$ cut. The false detection rate follows

$$\log_{10}(\mathcal{F}) = 2.67 - 0.97 \times {\rm SNR}. \qquad (7)$$

Equation 7 implies that caution should be taken when considering individual sources in the S2CLS catalogue at detection significance of less than $5\sigma$; follow-up confirmation and/or robust counterpart identification will be important for assessing the reality of sources

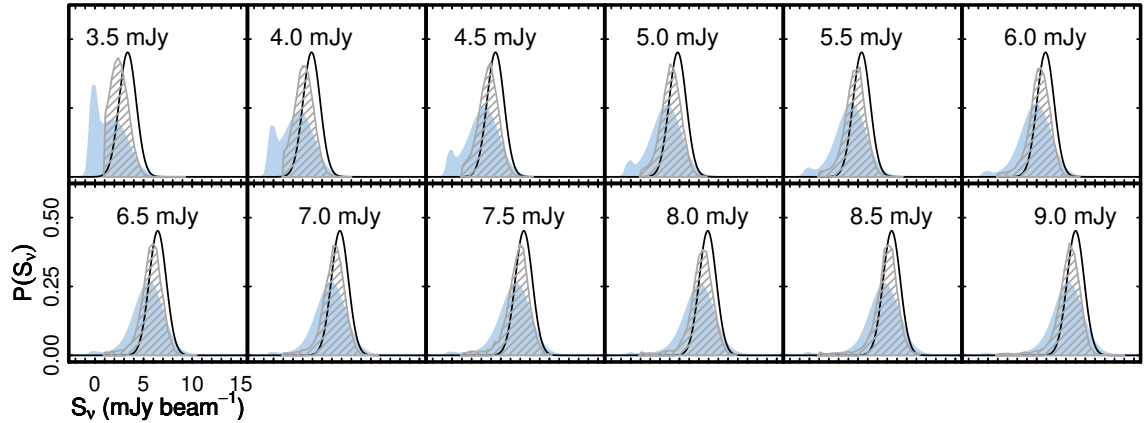

**Figure 9.** Comparison of deboosted flux density distributions for a Bayesian and empirical 'recovery' method (§2.5.1), using the UKIDSS-UDS field as an example. Both deboosting methods involve considering a model source distribution (down to a flux density of 1 mJy in this case). Each panel shows an observed flux probability distribution, assuming Gaussian uncertainties, for increasing observed flux. The solid and hatched distributions show the predicted intrinsic flux distribution for the Bayesian and direct methods respectively. In general the average boosting measured by the two methods agree well, converging as observed flux density increases, however the 'direct' method systematically predicts less boosting compared to the Bayesian approach; we discuss this in the main text.

detected close to the survey limit, and this work has already begun (e.g. Chen et al. 2016).

## 3 NUMBER COUNTS OF THE 850$\mu$M POPULATION

In Table 4 we present a sample of the S2CLS catalogue. The full catalogue contains 2,851 sources at a detection significance of $\geqslant 3.5\sigma$. The catalogue contains observed and deboosted flux densities, instrumental and deboosted flux density uncertainties, and individual completeness and false detection rates. The full catalogue and maps (match-filtered and non-match-filtered) are available at the DOI: http://dx.doi.org/10.5281/zenodo.57792

The surface density of sources per observed flux density interval $dN/dS$ – of a cosmological population is a simple measure of source abundance and a powerful tool for model comparisons. To measure the counts, for each catalogued source we first deboost the observed flux density using the empirical approach described in §2.5.1, and then apply the corresponding completeness correction for the deboosted (i.e. 'true') flux density. When deboosting, we consider the full intrinsic flux distribution as estimated by our simulation, accounting for the fact that a range of intrinsic flux densities can map onto an observed flux density. Therefore, we evaluate $dN/dS$ 1000 times; in each calculation every source is deboosted by randomly sampling the intrinsic flux distribution and completeness correcting each deboosted source accordingly. We take the mean of these 1000 realisations as the final number counts, with the standard deviation of $dN/dS$ in each bin as an additional uncertainty (to the Poisson error). We make a correction for each source based on the probability it is a false positive, using the empirical determination described in §2.5.3.

While the various corrections are intended to recover the 'true' underlying source distribution, it is important to confirm if any systematic biases remain, since the procedure for actually identifying sources is imperfect, as is the 'recovery' of injected model sources used to estimate flux boosting and completeness. To examine this we inject three different source count models into a jackknife noise map (of the UKIDSS-UDS field). One model is identical to the Schechter form used in §2.6.1 (Equation 3); in the other two models we simply adjust the faint end slope to $\gamma = 0.4$ and $\gamma = 2.4$, keeping the other parameters fixed. With knowledge of the exact model counts injected into the map, we can compare to the recovered counts before and after corrections have been applied. Figure 14 shows $([dN/dS]_{\rm rec} - [dN/dS]_{\rm true})/[dN/dS]_{\rm true}$ for the three models before and after corrections. In the absence of correction, flux boosting tends to result in the systematic overestimation of the number counts in all but the faintest flux bin, where incompleteness dominates, and the overestimation increases with increasing $\gamma$, as expected. After the corrections have been applied, there remains a slight underestimation in the counts in the faintest bin (3–4 mJy) at the 10% level, but in general the corrected 'observed' counts are in excellent agreement with the input model. The origin for the slight discrepancy is not clear, but it is likely that it simply stems from subtle effects not modelled well by our recovery simulation, and in particular what consititutes a 'recovered' source. One can observe a systematic effect that the $\gamma = 2.4$ and $\gamma = 0.4$ models are over- and under-estimated (respectively) at approximately the 10% level for the full observed flux range, but this is not a significant systematic uncertainty compared to shot noise expected from Poisson statistics. Given that the fiducial model we use in the completeness simulation is based on observed 850$\mu$m number counts, and the $\gamma = 2.4$ and $\gamma = 0.4$ models are rather extreme compared to empirical constraints, we consider this test as an adequate demonstration that our measured number counts are robust. Nevertheless, we apply a simple correction to the observed corrected counts by fitting a spline to the residual model counts in Figure 14 and apply this as a 'tweak' factor to the number counts on a bin-by-bin basis.

The S2CLS differential (and cumulative) number counts are presented in Table 4 and Figure 15. Tables of the number counts of individual fields are available in the electronic version of the paper. S2CLS covers a solid angle large enough to detect reasonable numbers of the rarer, bright sources at $S_{850} > 10$ mJy, allowing us to robustly measure the bright end of the observed 850$\mu$m number counts. As a guide, there are about ten sources with flux densities greater than 10 mJy per square degree. The 850$\mu$m source counts above 10 mJy clearly show an upturn in source density that is due to a mixture of local emitters and gravitationally lensed sources[2]. The wide area counts of *Herschel* demonstrated the same (predicted) phenomenon in the SPIRE bands (Negrello et al. 2010), and it has since been demonstrated that a simple bright submillimetre flux cut is highly effective at identifying strongly lensed sources once local galaxies have been rejected. The effect has already been observed in the millimetre regime: Vieira et al. (2010) detect the upturn in the 1.4 mm counts at $S_{1.4mm} > 10$ mJy from SPT over a 87 deg$^2$ survey, and Scott et al. (2012) have reported tentative evidence of an upturn in the counts at 1.1 mm at $S_{1.1mm} > 13$ mJy with AzTEC over 1.6 deg$^2$. Much larger 850$\mu$m surveys (exceeding 10 deg$^2$) could utlize a similar selection to cleanly identify lensed 850$\mu$m-selected high-redshift galaxies.

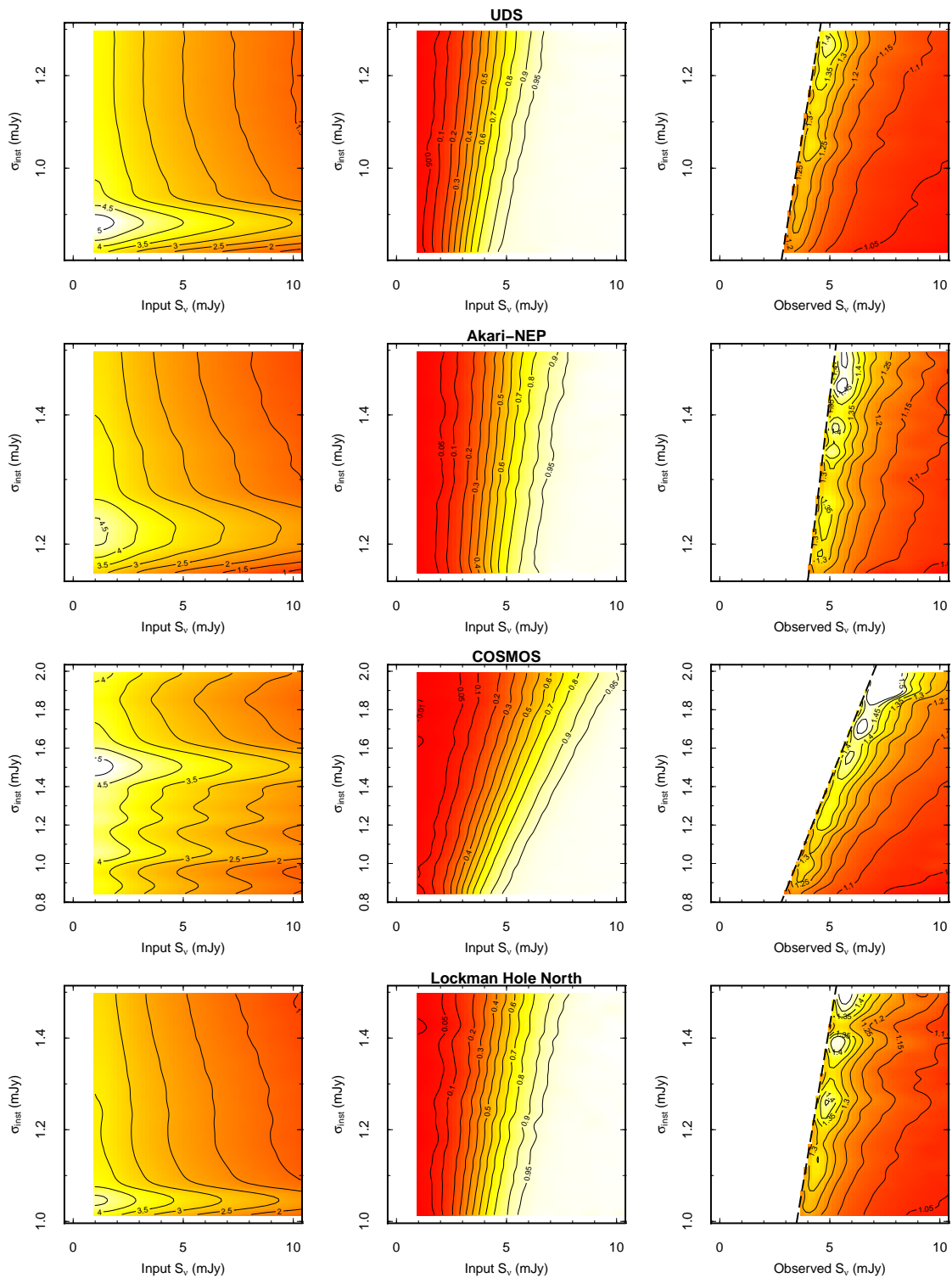

**Figure 10.** Two dimensional visualisations of the results of the recovery simulation in each field. The first column shows the number of artificial sources injected per bin of input flux density and local instrumental noise (labels are $\log_{10}(N)$). The prominent horizontal ridges clearly show the typical depth of the map. The middle column shows the completeness as a function of *true* flux density and local instrumental noise and the last column shows the average flux boosting as a function of *observed* flux density and local instrumental noise. The dashed line shows the $3.5\sigma$ detection limit.

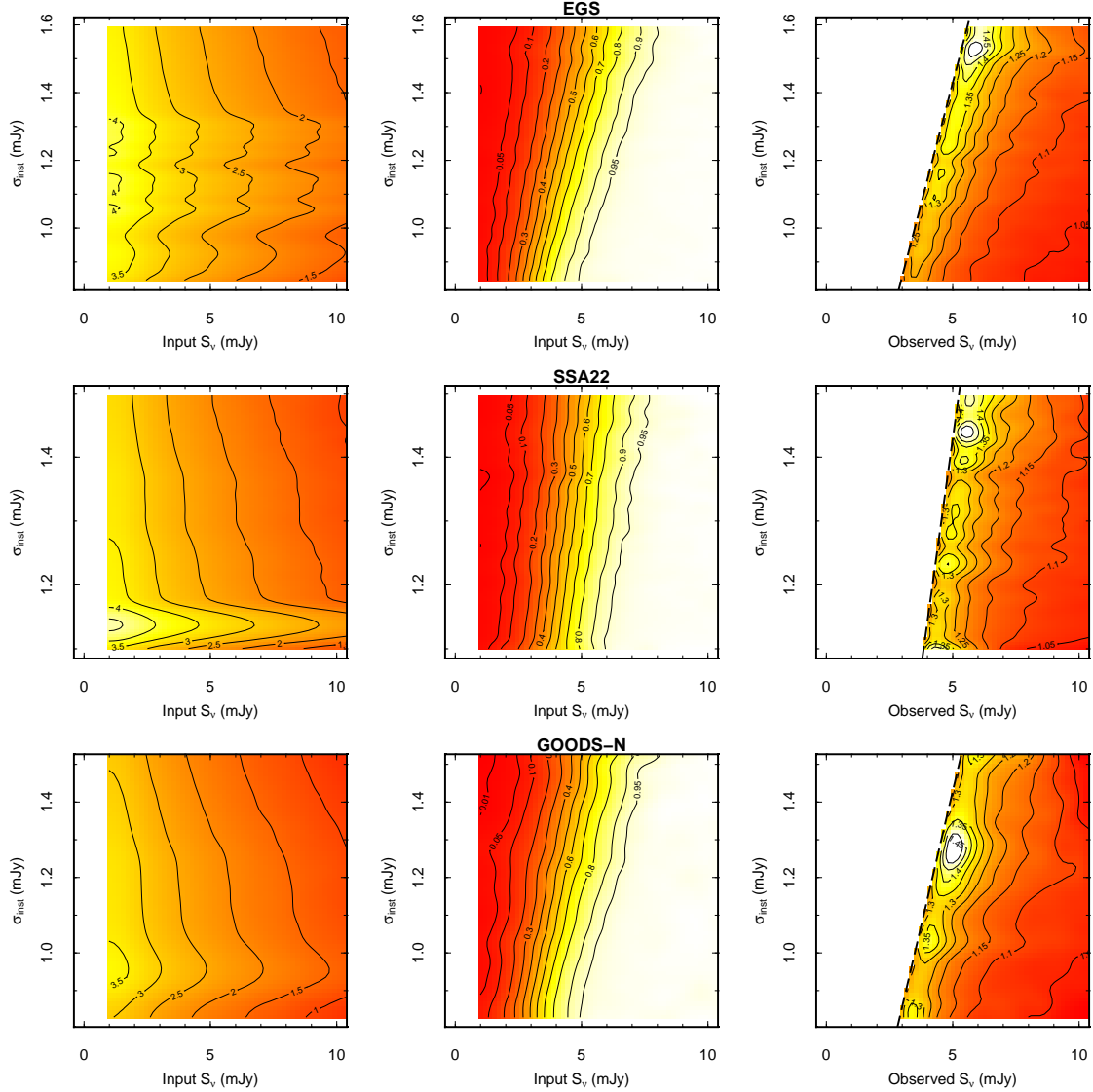

**Figure 10.** (continued)

The S2CLS number counts are in reasonable agreement with previous surveys for the flux range probed (for clarity, a non-exhaustive list of previous surveys, including recent SCUBA-2 results, are shown in Figure 15: Coppin et al 2006; Weiss et al. 2009; Casey et al. 2013; Chen et al. 2013), but with the large number of sources in S2CLS we can dramatically reduce the Poisson errors: in the faintest bin the Poisson uncertainty on the differential counts over the whole survey is just ~4%. We fit the combined differential counts (up to 20 mJy after which the local/lensing upturn starts to contribute significantly) with the Schechter functional form given in Equation 3. We find the best fit parameters $N_0 = 7180 \pm 1220 \,\mathrm{deg}^{-2}$, $S_0 = 2.5 \pm 0.4 \,\mathrm{mJy\,beam}^{-1}$ and $\gamma = 1.5 \pm 0.4$.

---

[2] Note that the Akari-NEP field contains the galactic object NGC 6543 (the Cat's Eye Nebula) - which is a ~200 mJy 850$\mu$m source

**Table 3.** Sample of the full S2CLS catalogue, listing the highest and lowest significance detections in each field. Coordinates are J2000, with the individual map astrometric solutions tied to the reference catalogue listed in Table 1. The $S_{850}^{\rm obs} \pm \sigma_{\rm inst}$ column gives the observed flux density and instrumental noise, $S/N$ gives the detection signal-to-noise ratio, and $S_{850} \pm \sigma_{\rm tot}$ gives the estimated true flux density and combined total (instrumental, deboosting, confusion) uncertainty. The final column $\log_{10}(\mathcal{F})$ is the logarithm of the false detection rate for the detection signal-to-noise ratio (Equation 7), negligible for bright sources, but important to consider for sources at the detection limit.

| S2CLS ID | Short name | R.A. | Dec. | $S_{850}^{\rm obs} \pm \sigma_{\rm inst}$ | $S/N$ | $S_{850} \pm \sigma_{\rm tot}$ | $\langle \mathcal{C} \rangle$ | $\log_{10}(\mathcal{F})$ |
|---|---|---|---|---|---|---|---|---|
| S2CLSJ175833+663757 | NEP.0001 | 17 58 33.60 | +66 37 57.7 | $195.4 \pm 1.2$ | 158.4 | $195.4 \pm 1.5$ | 1.00 | $-147.78$ |
| S2CLSJ175416+665117 | NEP.0329 | 17 54 16.57 | +66 51 17.0 | $4.3 \pm 1.2$ | 3.5 | $2.9 \pm 1.9$ | 0.25 | $-0.66$ |
| S2CLSJ100015+021548 | COS.0001 | 10 00 15.72 | +02 15 48.6 | $12.9 \pm 0.8$ | 15.2 | $12.9 \pm 1.2$ | 1.00 | $-11.78$ |
| S2CLSJ095936+022506 | COS.0733 | 09 59 36.09 | +02 25 06.5 | $5.4 \pm 1.5$ | 3.5 | $3.6 \pm 2.4$ | 0.25 | $-0.65$ |
| S2CLSJ141951+530044 | EGS.0001 | 14 19 51.56 | +53 00 44.8 | $16.3 \pm 1.2$ | 14.1 | $16.3 \pm 1.4$ | 1.00 | $-10.69$ |
| S2CLSJ141612+521316 | EGS.0227 | 14 16 12.05 | +52 13 16.8 | $3.7 \pm 1.0$ | 3.5 | $2.6 \pm 1.7$ | 0.28 | $-0.66$ |
| S2CLSJ123730+621258 | GDN.0001 | 12 37 30.73 | +62 12 58.5 | $12.8 \pm 1.0$ | 13.2 | $11.9 \pm 1.6$ | 1.00 | $-9.82$ |
| S2CLSJ123734+620736 | GDN.0068 | 12 37 34.33 | +62 07 36.5 | $5.0 \pm 1.4$ | 3.5 | $3.3 \pm 2.2$ | 0.23 | $-0.65$ |
| S2CLSJ104635+590748 | LHO.0001 | 10 46 35.78 | +59 07 48.0 | $12.0 \pm 1.0$ | 11.6 | $11.5 \pm 1.8$ | 0.99 | $-8.31$ |
| S2CLSJ104541+584640 | LHO.0219 | 10 45 41.47 | +58 46 40.0 | $4.1 \pm 1.2$ | 3.5 | $3.0 \pm 1.8$ | 0.29 | $-0.67$ |
| S2CLSJ221732+001740 | SSA.0001 | 22 17 32.50 | +00 17 40.4 | $14.5 \pm 1.1$ | 13.0 | $14.5 \pm 1.4$ | 0.96 | $-9.72$ |
| S2CLSJ221720+002024 | SSA.0198 | 22 17 20.23 | +00 20 24.4 | $3.9 \pm 1.1$ | 3.5 | $2.8 \pm 1.8$ | 0.28 | $-0.65$ |
| S2CLSJ021830-053130 | UDS.0001 | 02 18 30.77 | $-05$ 31 30.8 | $52.7 \pm 0.9$ | 56.7 | $52.7 \pm 1.2$ | 1.00 | $-51.18$ |
| S2CLSJ021823-051508 | UDS.1080 | 02 18 23.12 | $-05$ 15 08.9 | $3.1 \pm 0.9$ | 3.5 | $2.4 \pm 1.5$ | 0.30 | $-0.65$ |

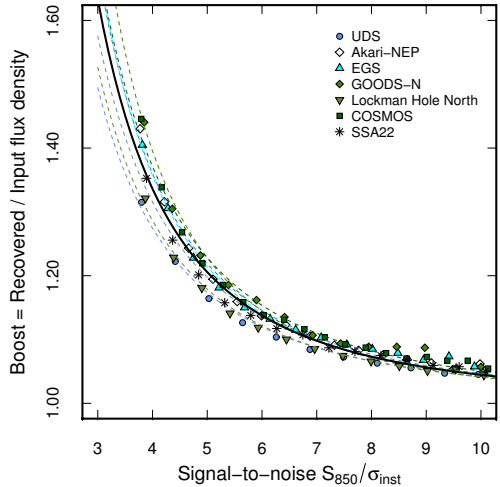

**Figure 11.** Average flux boosting as a function of signal-to-noise ratio, showing consistency at a fixed signal-to-noise level across different fields. The boosting can be described by a power law, however in practice we deboost sources individually based on their observed flux density and local instrumental noise, and drawing on the full probability distribution of true flux densities derived from our recovery simulation.

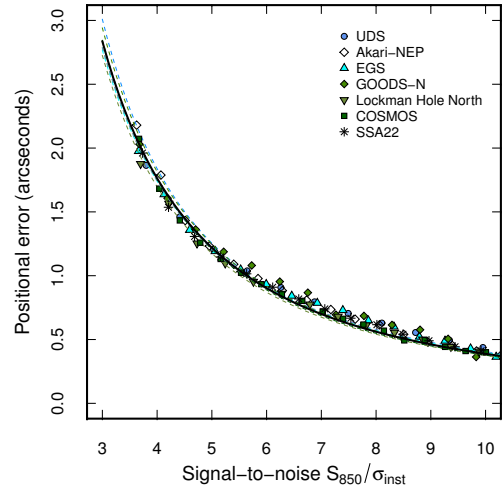

**Figure 12.** Average positional error based on the difference between input and recovered (peak) position from our recovery simulation. All fields follow a similar trend, with the positional uncertainty decreasing with increasing source significance. We fit the uncertainties with a simple power law to estimate the $1\sigma$ positional uncertainty as a function of observed signal-to-noise ratio (§2.5.2).

### 3.1 Field-to-field variance

Taking the full survey counts as an average measure of the abundance of submillimetre sources, with S2CLS we can now investigate field-to-field variance in the number counts in a consistent manner; this is important given that SMGs are thought to be a highly biased tracer of the matter field (Hickox et al. 2012; Chen et al. 2016). Letting $\rho(S) = N(> S)$, for each field we can consider the deviation of the counts compared to the mean density per flux bin: $\delta(S) = (\rho(S) - \langle \rho(S) \rangle)/\langle \rho(S) \rangle$. In Figure 16 we show the $\delta(S)$ measured for each field as a function of flux density, where uncertainties are the combined Poisson errors (obviously dominated by the single field counts). The field-to-field scatter on $\sim$0.5–1 degree scales is generally within 50% of the survey-averaged density and reasonably consistent with the Poisson errors. There are some hints that the GOODS-N field has a slightly elevated density compared to the mean (hints that were already apparent in the original SCUBA maps of this field, see Borys et al. 2002; Pope et al. 2005; Coppin et al. 2006), but this is marginal given the Poisson errors. However, to explore this further, and to quantify the significance of any overdensity, we can evaluate the field-to-field fluctuations on scales equivalent to the GOODS-N field taking into account cosmic variance.

Field-to-field variance in the observed number counts is caused by both shot noise and cosmic variance, with the latter defined as the excess variance in addition to Poisson noise (e.g. Somerville et al. 2004). We split the S2CLS survey (barring GOODS-N) into 16 independent fields of identical size to the GOODS-N field and count the number of sources in each field with deboosted flux densities greater than the 50% completeness limit in GOODS-N ($S_{850} \approx 4$ mJy). The mean number of sources is 18, with a standard deviation over the 16 fields of 5 sources, roughly consistent with the shot noise expected from Poisson statistics. The number of sources at the same limit in GOODS-N is $32 \pm 6$. It is clear from this simple analysis that the error budget on the counts is dominated by Poisson noise, but we can estimate what the expected contribution from cosmic variance is. Following the method of Trenti & Stiavelli (2008) which estimates the relative excess uncertainty in number counts due to cosmic variance in a flux limit survey, we find a contribution of 15–20% to the observed counts on scales of the GOODS-N field (note, we assume a Press-Schechter approach for the halo statistics). This assumes a mean redshift of $\langle z \rangle = 2.2$ and $\Delta z = 1$ and a wide range of halo filling factors, $f = 0.1$–1, corresponding to a mean bias of $b = 2.7$–4.3 for the SMG population. We can therefore quantify the significance of the tentative overdensity in GOODS-N as the difference in the number of sources in this field to the average over a region 16 times larger in an independent field (i.e. the rest of the S2CLS). We find $\Delta S(> 4 \text{mJy}) = 14 \pm 7$ taking into account Poisson noise and cosmic variance. Thus, the overdensity is signficant at only the $2\sigma$ level. GOODS-N is one of the most exhaustively studied extragalactic fields, and it is worth noting that overdensities of SMGs and star-forming galaxies have previously been reported here. For example, Daddi et al. (2009) report an overdensity of star-forming galaxies at $z \approx 4$, including SMGs, and Walter et al. (2012) report a $z \approx 5$ structure around the source HDF850.1, which happens to be one of the first SMGs to be identified (Hughes et al. 1998).

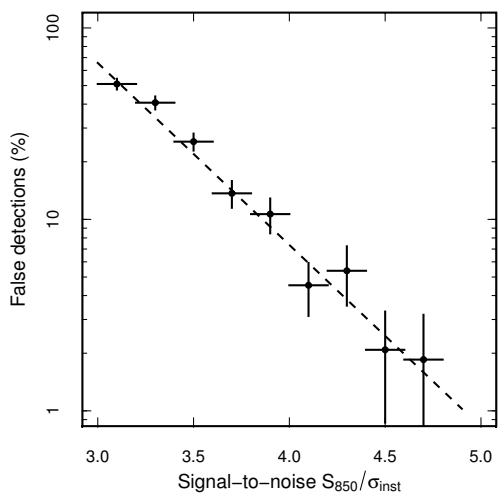

**Figure 13.** False detection rate averaged over the survey defined as the ratio of 'detections' in jackknife maps to real detections for sources at a fixed signal-to-noise limit. At our $3.5\sigma$ limit the false detection (or contamination) rate is 20%, falling to 6% at $4\sigma$ and is negligible for a $\geqslant 5\sigma$ cut. The implication is that, although the final S2CLS catalogue is cut at $3.5\sigma$, caution should be taken in the consideration of indivdual sources below a significance of $5\sigma$.

**Table 4.** Number counts measured in the full S2CLS. Flux density bins $\Delta S$ are 1 mJy wide. The flux density $S$ is the bin central and $S' = S - 0.5\Delta S$. Uncertainties on the counts are written such that the first set of errors are Poissonian and the second reflect the standard deviation of each bin of $dN/dS$ after 1000 realisations of the counts, where each source is deboosted (and completeness corrected) by sampling the deboosting probability distribution corresponding to the observed flux density and local noise. These uncertainties are of comparable magnitude to the Poisson errors.

| $S$ | $dN/dS$ | $N(>S')$ |
|---|---|---|
| (mJy) | (deg$^{-2}$ mJy$^{-1}$) | (deg$^{-2}$) |
| 3.5 | $451.0^{+17.1}_{-16.4} \pm 20.3$ | $1012.3^{+19.6}_{-19.2} \pm 19.6$ |
| 4.5 | $204.4^{+9.3}_{-8.9} \pm 8.8$ | $508.0^{+12.3}_{-12.0} \pm 9.7$ |
| 5.5 | $102.6^{+6.0}_{-5.7} \pm 5.1$ | $271.9^{+8.5}_{-8.2} \pm 6.5$ |
| 6.5 | $56.1^{+4.3}_{-4.0} \pm 3.8$ | $151.8^{+6.2}_{-6.0} \pm 4.3$ |
| 7.5 | $32.5^{+3.2}_{-2.9} \pm 2.5$ | $85.3^{+4.7}_{-4.4} \pm 3.1$ |
| 8.5 | $18.0^{+2.5}_{-2.2} \pm 2.0$ | $47.1^{+3.6}_{-3.3} \pm 2.3$ |
| 9.5 | $9.8^{+1.9}_{-1.6} \pm 1.4$ | $26.4^{+2.8}_{-2.5} \pm 1.6$ |
| 10.5 | $5.8^{+1.5}_{-1.2} \pm 1.0$ | $14.5^{+2.2}_{-1.9} \pm 1.2$ |
| 11.5 | $3.4^{+1.2}_{-0.9} \pm 0.8$ | $8.7^{+1.8}_{-1.5} \pm 0.8$ |
| 12.5 | $2.1^{+1.1}_{-0.7} \pm 0.6$ | $5.5^{+1.5}_{-1.2} \pm 0.6$ |
| 13.5 | $0.8^{+0.8}_{-0.4} \pm 0.4$ | $3.2^{+1.2}_{-0.9} \pm 0.5$ |
| 14.5 | $0.5^{+0.7}_{-0.3} \pm 0.3$ | $2.4^{+1.1}_{-0.8} \pm 0.3$ |
| 15.5 | $0.3^{+0.6}_{-0.2} \pm 0.1$ | $1.8^{+1.0}_{-0.7} \pm 0.2$ |

### 3.2 Comparison to models

At first it proved difficult for semi-analytic models of $\Lambda$CDM galaxy formation to reproduce the $850\mu$m number counts (Granato et al. 2000). The model of Baugh et al. (2005) provided a much better match to observed $850\mu$m (and Lyman-break Galaxy) number counts than previously achieved, but required a modification to the initial mass function (IMF) such that bursts of star formation have a more top heavy IMF than 'quiescent' star formation. While the motivation for this can be linked to astrophysical differences in the conditions of star formation in dense gas-rich starbursts (e.g. Padoan & Nordlund 2002), deviation from a universal IMF remains controversial. An additional problem was that the Baugh et al. model failed to predict the evolution of the *K*-band luminosity function. Recently Lacey et al. (2015) presented an update to the GALFORM model that adopts the best fitting $\Lambda$CDM cosmological parameters available from recent experiments, implementing more sophisticated treatments for star formation in disks, distinguishing molecular and atomic hydrogen (Lagos et al. 2011, 2012); dynamical friction timescales for mergers (Jiang et al. 2008) and stellar population synthesis models.

The Lacey et al. model counts are shown in Figure 15, and are in reasonable agreement with the data. This model still includes a mildly top heavy IMF (slope $x = 1$) for starbursts, without which it cannot reproduce the redshift distribution of $850\mu$m selected sources (Chapman et al. 2005; Simpson et al. 2014, see also Hayward et al. 2013). The model predicts a slightly elevated abundance of galaxies below the survey limit, however, an extrapolation of the Schechter fit to the S2CLS counts is in good agreement with the deeper observations of Chen et al. (2013). Nevertheless, the shape of the faint end slope is still to be properly determined empirically, which will most likely be through either a *P(D)* analysis of confused SCUBA-2 maps (e.g. Condon 1974; Pantanchon et al. 2009; Geach et al. in preparation), with the assistance of gravitational lensing (e.g. Knudsen et al. 2008; Chen et al. 2013) or through deep, unconfused ALMA surveys that can probe to the sub-mJy level, albeit over relatively small areas (e.g. Karim et al. 2013; Ono et al. 2014; Carniani et al. 2015; Oteo et al. 2016; Hatsukade et al. 2016; Dunlop et al. 2016). An important point to consider in comparing number counts to models is the issue of source blending and confusion in low resolution single-dish surveys. Therefore, we also show the results of Cowley et al. (2015), who take the same Lacey et al. (2015) GALFORM model, but predict the number counts after simulating observations with a single dish telescope with the same size beam as JCMT at $850\mu$m. Figure 15 shows that, over the observed flux density range, the beam-convolved predicted counts are consistent with the 'raw' model counts. The issue of 'multiplicity' of single-dish SMG detections has already started to be examined with the advent of sensitive interferometers (e.g. Hodge et al. 2013;

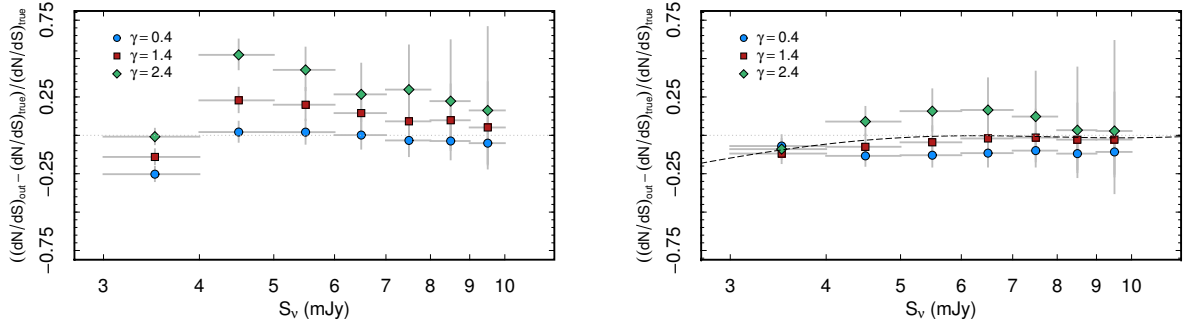

**Figure 14.** A comparison of recovered number counts to an ideal input model (Equation 3). Three input models are considered, differing only by the faint end slope $\gamma$: a series of fake catalogues are generated for each model by injecting sources into a jackknife map and then recovering them in a manner identical to the real data. In the left panel no deboosting, completeness or false positive correction has been applied to the recovered counts, showing the trend that steeper number counts are generally overpredicted (due to flux boosting) in all but the faint bin where incompleteness dominates. In the right panel the various corrections have been applied, illustrating that we can robustly recover the 'true' number counts, although there is still a slight (10%) underestimation of the counts in the faintest bin. The error bars in both panels reflect the Poisson uncertainties expected in a single field. The dashed line is a spline interpolation of the mean of the three models which we use as an additional tweak factor in measuring the number counts of the population. Interestingly, the two extreme count models we consider are, in general, systematically over and under predicted for the steeper and shallower faint end slope respectively; we discuss this in the main text.

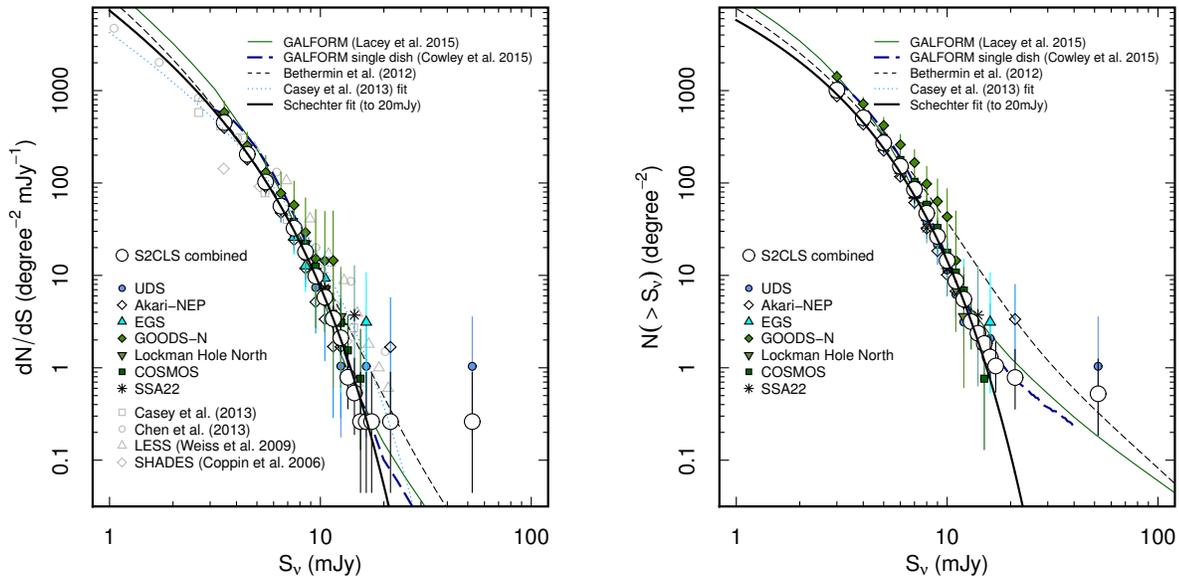

**Figure 15.** Number counts of $850\mu$m sources. The left panel shows the differential number counts for individual fields and the combined survey, along with a selection of data from the literature. Two model curves show the parametric evolving luminosity function model of Bethermin et al. (2012) and the semi-analytic (GALFORM) model of Lacey et al. (2015). The Cowley et al. (2015) line shows the same GALFORM model but taking into account source blending due to the $15''$ JCMT beam. The presence of foreground sources and the effect of gravitational lensing causes an upturn in the counts at bright flux densities at a level in reasonable agreement with the models (note that the GALFORM model does not include lensing) given the low number statistics at these bright flux densities. For clarity we only show error bars for the S2CLS data, which are evaluated from Poisson statistics (Gehrels et al. 1986). The right panel shows the cumulative counts where, for clarity, we only plot the S2CLS data, fit, and models.

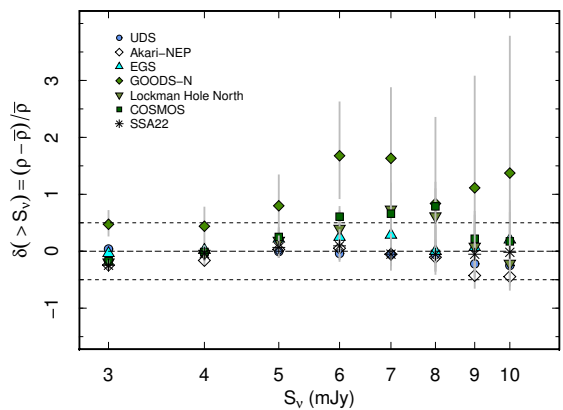

**Figure 16.** Field-to-field scatter in the integral number counts, relative to the mean density. The field-to-field scatter (on scales of 0.5–1°) across S2CLS is generally within 50% of the mean density, with the exception of GOODS-N, which has hints of an elevated density of SMGs compared to the mean, although this is marginal with the Poisson uncertainties. We discuss this in §3.1.

Simpson et al. 2015), and it is important to stress that comparisons of source abundances (between both models and data) should adopt a consistent reference resolution.

While the semi-analytic models aim to simultaneously reproduce all the main 'bulk' observational tracers of the galaxy population over cosmic time (i.e. the mass function, luminosity functions, number counts, clustering, etc.) in a single framework, an alternative approach to predicting the submillimetre number counts is through phenomenological modelling. Bethermin et al. (2012) present a model that considers the evolution of the space density of so called 'main sequence' (i.e. normal) star-forming galaxies and luminous starbursts, fitting parametric models (with assumptions about the underlying galaxy SEDs) to observed number counts across the infrared, submillimetre and radio bands. We show the Bethermin et al. model (including the strong lensing contribution) for the SCUBA-2 850$\mu$m band in Figure 15. Again, this is in reasonable agreement with the observations over the flux range probed by the observations. The new 850$\mu$m number counts presented here could be used to provide improved fits to phenomenological models such as this.

## 4 SUMMARY

We have presented the 850$\mu$m maps and catalogues of the James Clerk Maxwell Telescope SCUBA-2 Cosmology Legacy Survey, the largest of the JCMT Legacy Surveys, completed in early 2015. With hundreds of hours of integration time in reasonable submillimetre observing conditions (zenith opacity $\tau_{225\text{GHz}} = 0.05$–0.1) S2CLS has mapped seven well-known extragalactic survey fields: UKIDSS-UDS, *Akari*-NEP, COSMOS, GOODS-N, Extended Groth Strip, Lockman Hole North and SSA22. The total scientifically useful survey area is approximately 5 deg$^2$ at a sensitivity of under 2 mJy beam$^{-1}$, with a median depth per field of approximately 1.2 mJy beam$^{-1}$, approaching the confusion limit (which we have determined is approximately $\sigma_c \approx 0.8$ mJy beam$^{-1}$). This is by far the largest and deepest survey of submillimetre galaxies yet undertaken in this waveband and provides a rich legacy data source. We have detected nearly 3,000 submillimetre sources at the $\geqslant 3.5\sigma$ level, an order of magnitude increase in the number of catalogued 850$\mu$m-selected sources to date.

In this work we have used the S2CLS catalogue to accurately measure the number counts of submillimetre sources, dramatically reducing Poisson errors and allowing us to investigate field-to-field variance. The wide nature of the survey makes it possible to detect large numbers of bright (>10 mJy), but rare ($\sim$10 per square degree), submillimetre sources, and we observe the distinctive upturn in the number counts caused by strong gravitational lensing of high redshift galaxies and a contribution from local sources of submillimetre emission. The S2CLS catalogue and maps offer a route to a tremendous range of follow-up work, both in pin-pointed multi-wavelength identification and follow-up of the catalogued sources (e.g. Simpson et al. 2015ab, Chen et al. 2016) and in statistical analyses of the catalogues and pixel data. Cross-correlation of the submillimetre maps and galaxy catalogues is already proving a treasure trove of discovery, linking UV/optical/near-infrared-selected samples to submillimetre emission (Banerji et al. 2015; Coppin et al. 2015; Smith et al. 2016 in preparation; Bourne et al. in preparation). The S2CLS survey subtends an area equivalent to over $10^5$ times the ALMA primary beam at 850$\mu$m, and the synergy between large area single dish surveys such as S2CLS, and the detailed interferometric follow-up possible with ALMA (and other sensitive (sub)mm interferometers) is clear. High-resolution interferometric follow-up in the submillimetre has already proven efficient and fruitful, with ALMA and SMA imaging of the brightest (>9 mJy) sources revealing a complex morphological mix, allowing us to investigate the true nature of the SMGs identified in large beam single dish surveys (Simpson et al. 2015, Chapman et al. in preparation). We release the 3.5$\sigma$-cut catalogue of all S2CLS sources as part of this publication, along with the 850$\mu$m maps for exploitation by the community. The data are available at the DOI http://dx.doi.org/10.5281/zenodo.57792

## ACKNOWLEDGEMENTS

The authors thank M. Bethermin for supplying the SCUBA-2 850$\mu$m number count predictions. The James Clerk Maxwell Telescope is now operated by the East Asian Observatory on behalf of The National Astronomical Observatory of Japan, Academia Sinica Institute of Astronomy and Astrophysics, the Korea Astronomy and Space Science Institute, the National Astronomical Observatories of China and the Chinese Academy of Sciences (Grant No. XDB09000000), with additional funding support from the Science and Technology Facilities Council of the United Kingdom and participating universities in the United Kingdom and Canada. The data presented in this paper was taken as part of Program ID MJLSC02. It is a pleasure to thank the entire staff of the JCMT for their superb support throughout the S2CLS campaign. Special thanks is due to Iain Coulson, Jessica Dempsey, Jim Hoge, Harriet Parsons, Callie Matulonis, William Montgomerie and Holly Thomas. This research used the facilities of the Canadian Astronomy Data Centre operated by the National Research Council of Canada with the support of the Canadian Space Agency.

This work used the DiRAC Data Centric system at Durham University, operated by the Institute for Computational Cosmology on behalf of the STFC DiRAC HPC Facility (www.dirac.ac.uk). This equipment was funded by BIS National E-infrastructure capital grant ST/K00042X/1, STFC capital grant ST/H008519/1, and STFC DiRAC Operations grant ST/K003267/1 and Durham University. DiRAC is part of the National E-Infrastructure.

JEG is supported by a Royal Society University Research Fellowship. IRS, AMS, JMS, ALRD, DMA, ACE and CGL acknowledge support from ST/L00075X/1. IRS also acknowledges support from the ERC Advanced Grant DUSTYGAL (321334) and a Royal Society Wolfson Merit Award. JSD acknowledges the support of the European Research Council through the award of an Advanced Grant. RJI acknowledges support from ERC in the form of the Advanced Investigator Programme COSMICISM (321302). AK acknowledges support by the Collaborative Research Council 956, sub-project A1, funded by the Deutsche Forschungsgemeinschaft (DFG). KK acknowledges supports from the Swedish Research Council.

[2] *Institute for Astronomy, University of Edinburgh, Royal Observatory, Blackford Hill, Edinburgh, EH9 3HJ*
[3] *Department of Physics & Astronomy, University of British Columbia, 6224 Agricultural Road, Vancouver, BC, V6T 1Z1, Canada*
[4] *Centre for Extragalactic Astronomy, Department of Physics, Durham University, South Road, Durham, DH1 3LE*
[5] *Leiden Observatory, Leiden University, P.O. box 9513, 2300 RA Leiden, The Netherlands*
[6] *School of Physics and Astronomy, University of Nottingham, University Park, Nottingham, NG9 2RD*
[7] *Instituto Nacional de Astrofísica Óptica y Electrónica, Calle Luis Enrique Erro No. 1, Sta. Ma. Tonantzintla, Puebla, México*
[8] *European Southern Observatory, Karl-Schwarzschild-Str. 2, D-85748 Garching, Germany*
[9] *Institute of Astronomy, University of Cambridge, Madingley Road, Cambridge, CB3 OHA*
[10] *Department of Astronomy & Physics, St Mary's University, 923 Robie St., Halifax, Nova Scotia, B3H 3C3, Canada*
[11] *Department of Physics & Astronomy, University of Leicester, University Road, Leicester, LE1 7RH*
[12] *H. H. Wills Physics Laboratory, University of Bristol, Tyndall Avenue, Bristol, BS8 1TL*
[13] *Herzberg Astronomy and Astrophysics, National Research Council Canada, 5071 West Saanich Road, Victoria, BC V9E 2E7, Canada*
[14] *Department of Physics and Atmospheric Science, Dalhousie University Halifax, NS, B3H 3J5, Canada*
[15] *SKA Headquarters, The University of Manchester, Manchester, M13 9PL*
[16] *Astronomy Centre, Department of Physics and Astronomy, University of Sussex, Brighton, BN1 9QH*
[17] *Astrophysics Group, Imperial College London, Blackett Laboratory, Prince Consort Road, London, SW7 2AZ*
[18] *Institute for Computational Cosmology, Department of Physics, Durham University, South Road, Durham, DH1 3LE*
[19] *School of Physics and Astronomy, Cardiff University, Queen's Buildings, The Parade, Cardiff, CF24 3AA*
[20] *Virginia Polytechnic Institute & State University Department of Physics, MC 0435, 910 Drillfield Drive, Blacksburg, VA 24061, USA*
[21] *Department of Physics, University of Oxford, Keble Road, Oxford, OX1 3RH*
[22] *Department of Physics, University of the Western Cape, Bellville 7535, South Africa*
[23] *Department of Earth and Space Science, Chalmers University of Technology, Onsala Space Observatory, SE-43992 Onsala, Sweden*
[24] *Argelander-Institut für Astronomie, Universität Bonn, Auf dem Hügel 71, D-53121 Bonn, Germany*
[25] *Kapteyn Institute, University of Groningen, P.O. Box 800, 9700 AV Groningen, The Netherlands*
[26] *Mullard Space Science Laboratory, University College London, Holmbury St Mary Dorking, Surrey, RH5 6NT*
[27] *Space Science & Technology Department, Rutherford Appleton Laboratory, Chilton, Didcot, Oxfordshire, OX11 0QX*
[28] *UK Astronomy Technology Centre, Royal Observatory, Blackford Hill, Edinburgh, EH9 3HJ*
[29] *Robert Hooke Building, Department of Physical Sciences, The Open University, Milton Keynes, MK7 6AA*
[30] *Gemini Observatory, Northern Operations Center, 760 N. A'ohoku Place, Hilo, HI 96720, USA*
[31] *Department of Physics and Astronomy, Sonoma State University, 1801 East Cotati Avenue, Rohnert Park, CA 94928-3609, USA*
[32] *Department of Physics, McGill University, 3600 rue University, Montreal, QC H3A 2T8, Canada*
[33] *Canadian Astronomy Data Centre, National Research Council Canada, 5071 West Saanich Road, Victoria, BC, V9E 2E7, Canada*
[34] *Center for Detectors, School of Physics and Astronomy, Rochester Institute of Technology, Rochester, NY 14623, USA*
[35] *Jet Propulsion Laboratory, 4800 Oak Grove Drive, Pasadena, CA 91109, USA*